\documentclass[11pt]{article}
\usepackage[margin=1in]{geometry}
\usepackage{graphicx,url,wrapfig}
\usepackage[british]{babel}

\def\beq{\begin{equation}}
\def\eeq#1{\label{#1}\end{equation}}
\def\eeqn{\end{equation}}

\def\beqa{\begin{eqnarray}}
\def\eeqa#1{\label{#1}\end{eqnarray}}
\def\eeqan{\end{eqnarray}}

\let\bar=\overbar

\def\Dslash{\not{\hbox{\kern-4pt $D$}}}
\def\dslash{\not{\hbox{\kern-2pt $\del$}}}

\def\msb{{\bar{\ssstyle M \kern -1pt S}}}

\def\Title#1{\begin{center} {\Large {\bf #1} } \end{center}}
\def\Author#1{\begin{center} {\normalsize {\sc #1} } \end{center}}
\def\Institution#1{\begin{center} {\normalsize {\it #1} } \end{center}}
\def\Abstract#1{\noindent {\normalsize {\bf Abstract:} {\normalfont #1}}}
\def\Conference{\vspace{4mm}\begin{raggedright} {\normalsize {\it Talk presented at the 2019 Meeting of the Division of Particles and Fields of the American Physical Society (DPF2019), July 29--August 2, 2019, Northeastern University, Boston, C1907293.} } \end{raggedright}\vspace{4mm}}

\begin{document}

\Title{Observation of Complex Time Structures in the Cosmic Ray Fluxes by the Alpha Magnetic Spectrometer on the ISS}

\Author{Davide Rozza for the AMS-02 Collaboration}

\Institution{INFN and University of Milano-Bicocca, Piazza della Scienza 3, 20125 Milano, Italy}

\Abstract{We present high-statistics, precision measurements by AMS of the detailed time and rigidity dependence of the primary cosmic-ray electron, positron, proton and helium fluxes over 79 Bartels rotations from May 2011 to May 2017 in the energy range from 1 to 50 GeV. For the first time, the charge-sign dependent modulation during solar maximum has been investigated in detail by leptons alone. We report the observation of short-term structures on the timescale of months coincident in all the fluxes. These structures are not visible in the positron-to-electron flux ratio. The precision measurements across the solar polarity reversal show that the ratio exhibits a smooth transition over $\sim$800 days from one value to another.}

\Conference 
\section{Introduction}
Interstellar cosmic rays (CR) propagate inside the Galaxy interacting with other particles, magnetic and radiation fields. When these particles enter into the heliosphere, the solar modulation process (diffusion, convection, particle drift and adiabatic energy loss) reduces the differential intensity for energy below tens of GeV. This quantity is not stable, but changes according to the temporal evolution of the Sun. Our star reverses its magnetic field every 11 years (22 years for a complete cycle). This cycle can be observed in the time evolution of the sunspot number. During the solar magnetic field reversal, the sunspot number increases (high solar activity period), while it decreases when the solar magnetic field polarity is well defined (low solar activity period). Strong emission of plasma can occur during high solar activity phases, e.g. during coronal mass ejection (CME) whose effects can last from days to months.\\
Time structures in cosmic rays can be observed with detectors in different environments. Alpha Magnetic Spectrometer (AMS) is the last generation of particle spectrometer operating in space on board of the International Space Station (ISS). The wide energy range, the large acceptance and the long exposure time allow AMS to perform accurate measurements of the cosmic ray fluxes as a function of energy and time. In this work, we present the time evolution of the proton, helium, electron and positron fluxes observed by the same detector (AMS) and over the same time period, from May 2011 up to May 2017. These precise measurements are providing constraints on solar modulation models \cite{Corti,Boschini} as well as dark matter model \cite{Fornengo,Cirelli,Yuan} predictions. In addition the study of time variation of cosmic rays can be useful to plan the next generation of manned space missions.
\newpage
\section{AMS-02 detector}
\begin{wrapfigure}{l}{0.4\textwidth}
\label{fig:AMS}
\includegraphics[width=0.4\textwidth]{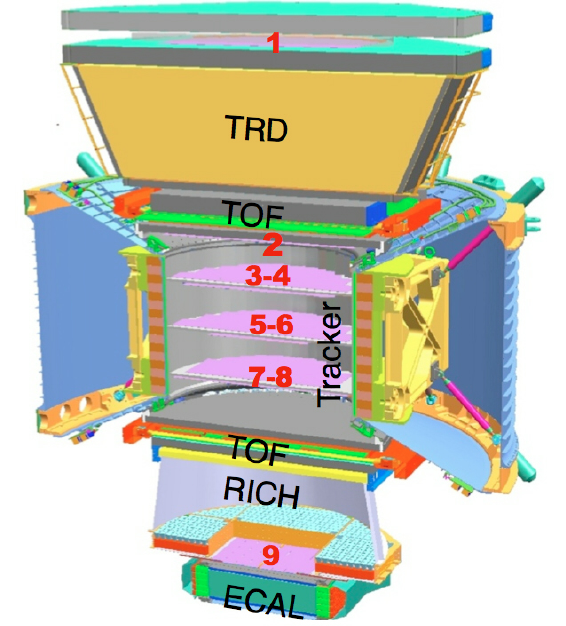}
\caption{AMS-02 apparatus.}
\end{wrapfigure}
AMS-02 is a large acceptance multi-purpose spectrometer. This particle detector was installed on the International Space Station in May 2011. In eight years, more than 140 billion of charged cosmic rays have been detected. The objectives of the experiment are: pri\-mor\-di\-al antimatter search with sensitivity of $10^{-9}$, dark matter search in antimatter channels (e.g. anti-protons, positrons, anti-deuterium), cosmic ray composition and spectral analysis. The layout and description of the AMS detector are presented in Ref.~\cite{AMS}. AMS is composed of four planes of time-of-flight (TOF) counters, a transition radiation detector (TRD), a ring imaging \v{C}erenkov detector (RICH), an electromagnetic calorimeter (ECAL), a permanent magnet and nine layers of silicon tracker. Six tracker layers are located inside the magnet to reconstruct the bending of the cosmic ray trajectories inside the detector allowing to measure both rigidity and charge sign. The particle rigidity ($R$) is evaluated as $R=p/Z$, combining information of the particle momentum ($p$) and charge ($Z$). Cosmic ray charge is measured independently by seven instruments. The calorimeter gives information related to the energy of electrons and positrons crossing it. TRD and ECAL are used to separate electrons and positrons from protons and antiprotons. The efficiency of the detector is evaluated by mean of Monte Carlo simulated events that were produced using a dedicated program developed by the collaboration based on the GEANT-4.10.1 package~\cite{MC}. The program simulates electromagnetic and hadronic interactions of particles in the material of the AMS and generates detector responses. The Monte Carlo event samples have enough statistics such that they do not contribute to the errors.

\section{Data analysis}
The time evolution of the proton flux from 1 to 60 GV is based on $846\times10^{6}$ events, the helium flux from 1.9 to 60 GV includes $112\times10^{6}$ events, while the electron and positron fluxes from 1 to 50 GeV are based on $23.5\times10^{6}$ particles. Fluxes were evaluated with a time binning of one Bartels rotation (i.e. 27 days) from May 2011 to May 2017. The flux ($\Phi_i$) of the $i^{th}$ Bartels rotation is:
\begin{equation}\label{EqFlux}
 \Phi_i=\frac{N_i}{A_{eff,i}T_i\epsilon\Delta R}
\end{equation}
defined by the ratio of the number of primary cosmic rays ($N_i$) over the exposure time ($T_i$), the effective acceptance ($A_{eff,i}$), the trigger efficiency ($\epsilon$) and the rigidity bin width ($\Delta R$ or energy bin width $\Delta E$ for electrons and positrons). The time collected in these analysis includes seconds during which AMS was in normal operating conditions, the detector was pointing within 40$^{\circ}$ of the local zenith and the ISS was outside the South Atlantic Anomaly region. AMS-02, located inside the magnetosphere, detects particles in different geomagnetic regions where a different geomagnetic cut-off is present. The geomagnetic cut-off is the rigidity limit above which only primary cosmic rays, coming from outside the magnetosphere, are allowed to reach a point inside the magnetosphere, while below it, only secondary particles produced in the atmosphere or trapped in the magnetic field lines are present. The cut-off reaches few tens of GV at the equator and decreases below 1 GV towards the geomagnetic poles. Therefore, particles with rigidity of 1 GV can be seen only close to these poles. Thus, the exposure time ($T_i$) of equation (\ref{EqFlux}) is determined as a function of the rigidity (or energy) by counting the number of livetime-weighted seconds at each location above the geomagnetic cut-off. Proton and helium events were selected as described in Refs.~\cite{AMSp,AMSHe,AMSpHet}, while electrons and positrons were analyzed as in Refs.~\cite{AMSep,AMSept}.

\section{Results}
\begin{figure}[htb]
\centering
\includegraphics[width=1\textwidth]{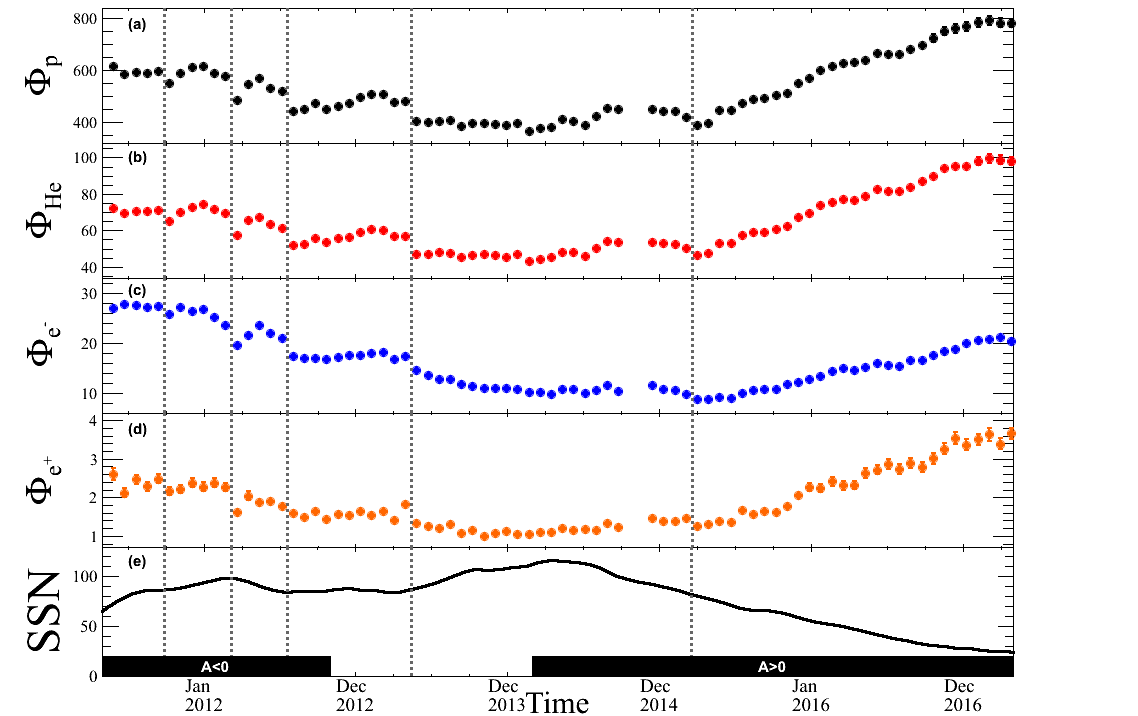}
\caption{Comparison of the time structures in Cosmic Rays observed by AMS-02: in panel (a) proton flux $\Phi_p [GV\ s\ sr\ m^2]^{-1}$ in the rigidity bin [1.92-2.15] GV; in panel (b) helium flux $\Phi_{He} [GV\ s\ sr\ m^2]^{-1}$ in the rigidity bin [1.92-2.15] GV; in panel (c) electron flux $\Phi_{e^{-}} [GeV\ s\ sr\ m^2]^{-1}$ in the energy bin [1.01-1.22] GeV; in panel (d) positron flux $\Phi_{e^{+}} [GeV\ s\ sr\ m^2]^{-1}$ in the energy bin [1.01-1.22] GeV; in panel (e) 13-month smoothed monthly total sunspot number (SSN) \cite{SSN} from May 2011 to May 2017. Error bars are combination of statistics and total systematic uncertainties. Vertical dashed lines denote structures observed in all AMS particle channels. The two black regions at the bottom identify the polarity of the solar magnetic field and the period of field-reversal (from November 2012 to March 2014).}
\label{fig:pHeeleposSSN}
\end{figure}

The proton, helium, electron and positron fluxes are tabulated in the Ref.~\cite{SMpHet,SMept}. In Figure~\ref{fig:pHeeleposSSN}, the proton and the helium fluxes at rigidity $[1.92-2.15]$ GV are shown in panels (a) and (b) respectively. Electron and positron fluxes at energy $[1.01-1.22]$ GeV are reported in panels (c) and (d) of Figure~\ref{fig:pHeeleposSSN}.
Error bars take into account statistics and the total systematics. Contributions to the total systematic error come from: acceptance, background contamination, geomagnetic cutoff, event selection, unfolding, rigidity resolution function, absolute rigidity scale and time dependent systematic errors coming from the trigger and the reconstruction efficiency. Proton and helium fluxes exhibit structures as a function of time at low rigidities disappearing with increasing rigidity. These structures are nearly identical in both time and relative amplitude. Above 40 GV, both fluxes are time independent. Prominent and distinct time structures are visible in both $e^-$-flux and $e^+$-flux at energies up to 20 GeV, as well as in $p$-flux and $He$-flux up to 20 GV, while no evidence of time dependent structures is exhibited at higher energy. For a qualitative comparison, 13-month smoothed monthly total sunspot number \cite{SSN} is reported in panel (e) of Figure~\ref{fig:pHeeleposSSN} to indicate the solar activity variation during time period 2011-2017. The two black regions identify the two polarity of the solar magnetic field and the time region characterized by the magnetic field reversal of the Sun (from November 2012 to March 2014). Vertical dashed lines denote structures observed in all AMS flux measurements. These lines correspond to short term solar events, like coronal mass ejection (CME), impacting on Earth. As representative example, the second corresponds to March 2012, where a CME caused a strong Forbush decrease \cite{March2012,AMS2012,Bindi2015}. After the last vertical dashed line, corresponding to April 2015, structures in the proton and helium fluxes are considerably reduced and all fluxes increase.\\
\begin{figure}[htb]
\centering
\includegraphics[width=1\textwidth]{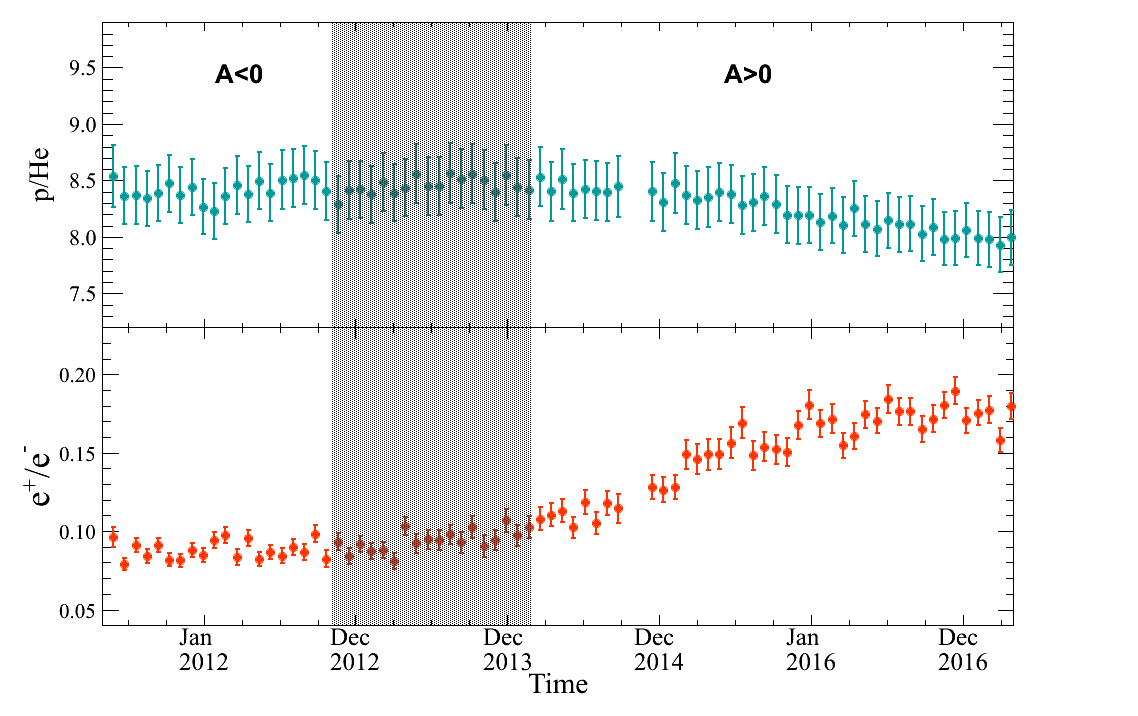}
\caption{The AMS $p/He$ flux ratio for the rigidity bin $[1.92-2.15]$ GV (top) and the AMS $e^+/e^-$ flux ratio for the energy bin $[1.01-1.22]$ GeV (bottom) as function of time from May 2011 to May 2017. Error bars are combination of statistics and total systematic uncertainties. The three time regions identify the different solar magnetic field polarities ($A<0$, $A>0$) and the period characterized by the magnetic field reversal of the Sun (from November 2012 to March 2014).}
\label{fig:ProHePosEle}
\end{figure}
More information regarding the cosmic ray propagation in the heliosphere come from study of the ratios between particle fluxes. In Figure~\ref{fig:ProHePosEle}, the AMS $p/He$ flux ratio for the rigidity bin $[1.92-2.15]$ GV (top) and the AMS $e^+/e^-$ flux ratio for the energy bin $[1.01-1.22]$ GeV (bottom) as a function of time from May 2011 to May 2017 are reported. The three time regions identify the solar magnetic field polarity ($A<0$), the reversal (from November 2012 to March 2014) and the opposite polarity ($A>0$) of the Sun. The AMS $p/He$ flux ratio is characterized by two different behaviors: it is time independent above 3 GV, while it has a long term decrease time dependence below 3 GV. The long-term decrease appears around February 2015 corresponding to the last dashed line reported in Figure~\ref{fig:pHeeleposSSN}. Interpretation can be addressed to: differences in the diffusion coefficient for the two species, differences in the $p$ and $He$ local interstellar spectra, differences in the $^3He$ and $^4He$ isotopic composition \cite{Corti,Boschini,Cholis,Gieseler, Tomassetti,Vos,Gloeckler,Cummings,HelMod,Toma2}.\\
The AMS $e^+/e^-$ flux ratio gives information related to the charge sign dependence related to the solar activity (see e.g. \cite{DellaTorre}). In this ratio, short term variations in the fluxes largely cancel out and a long term trend appears. Below 20 GeV, the ratio can be described by a smooth transition from one constant value to an higher one \cite{AMSept}. The duration of the transition is energy independent between 1 and 6 GeV and corresponds to a constant value of 830$\pm$30 days. The amplitude of this transition is energy dependent and is compatible with 0 above 20 GeV, while the midpoint of the transition shifts by 260$\pm$30 days from 1 to 6 GeV.

\section{Conclusions}
For the first time, proton, helium, electron and positron fluxes are simultaneously measured with the same precise detector for an extended period of time. All species exhibit large time variation at low rigidities (energies) which decrease with increasing rigidity (energy).\\
The accurate AMS-02 data provide precious information for the development of refined solar modulation models \cite{Boschini,HelMod}. These measurements will continue for the entire AMS-02 mission up to 2024, covering the descending phase of the current field polarity. In addition, the time dependence of several more cosmic ray species will be provided. Questions regarding the diffusion mechanisms over a solar cycle and the charge sign dependence of the solar modulation will be addressed by the new AMS data.


\end{document}